\documentclass{article}
\usepackage{emulateapj,apjfonts}
\usepackage{lscape}

\newcommand{\etal}{et~al.\ }

\def\ts{\thinspace}

\newdimen\sa  \def\sd{\sa=.1em \ifmmode $\rlap{.}$''$\kern -\sa$
                               \else \rlap{.}$''$\kern -\sa\fi}

\reversemarginpar

\begin{document}

\lefthead{Supermassive Black Holes and Galaxy Cores}

\righthead{Kormendy \& Bender}

\centerline{\null}\vskip -50pt

\title{Correlations Between Supermassive Black Holes, Velocity Dispersions, and\\
       Mass Deficits in Elliptical Galaxies with Cores\altaffilmark{1}}

\author{
John Kormendy\altaffilmark{2} and
Ralf Bender\altaffilmark{3,4}
}

\altaffiltext{1}{Based on observations made with the NASA/ESA {\it Hubble Space Telescope}, 
                 obtained from the Data Archive at STScI, which is operated by AURA, Inc., under 
                 NASA contract NAS 5-26555.  These observations are associated with program
                 numbers 5477, 7868, 8686, and 9401.} 

%\altaffiltext{2}{Based on observations obtained at the Canada-France-Hawaii 
%                 Telescope (CFHT), which is operated by the National Research 
%                 Council of Canada, the Institut National des Sciences de l'Univers
%                 of the Centre National de la Recherche Scientifique of France, 
%                 and the University of Hawaii.} 

\altaffiltext{2}{Department of Astronomy, University of Texas at Austin,
                 1 University Station C1400, Austin,
                 Texas 78712-0259; kormendy@astro.as.utexas.edu}

\altaffiltext{3}{Universit\"ats-Sternwarte, Scheinerstrasse 1,
                 M\"unchen D-81679, Germany}

\altaffiltext{4}{Max-Planck-Institut f\"ur Extraterrestrische Physik,
                 Giessenbachstrasse, D-85748 Garching-bei-M\"unchen, Germany; 
                 bender@mpe.mpg.de}

\pretolerance=15000  \tolerance=15000

\begin{abstract} 

      High-dynamic-range surface photometry in a companion paper makes possible accurate measurement 
of the stellar light deficits $L_{\rm def}$ and mass deficits $M_{\rm def}$ associated with the cores 
of elliptical galaxies.  We show that $L_{\rm def}$ correlates with the velocity dispersion $\sigma$ 
of the host galaxy bulge averaged outside the central region that may be affected by a 
supermassive black hole (BH).  We confirm that $L_{\rm def}$ correlates with BH mass
$M_\bullet$.  Also, the fractional light deficit $L_{\rm def}/L_*$ correlates with $M_\bullet/M_*$, 
the ratio of BH mass to the galaxy stellar mass.  All three correlations have scatter similar to
or smaller than the scatter in the well known correlation between $M_\bullet$ and $\sigma$.  
The new correlations are remarkable in view of the dichotomy between ellipticals with cores and 
those with central extra light.  Core light deficit correlates closely with $M_\bullet$ and $\sigma$, 
but extra light does not.  This supports the suggestion that extra light Es are made in wet mergers
with starbursts whereas core Es are made in dry mergers.  After dry mergers, cores are 
believed to be scoured by BH binaries that fling stars away as their orbits decay or by BHs that
sink back to the center after recoiling from anisotropic gravitational radiation emitted when they merge.  
Direct evidence has been elusive.  We interpret the new correlations as the ``smoking gun'' that 
connects cores with BHs.  Together, the $M_\bullet$ -- $\sigma$ and $M_\bullet$ -- $L_{\rm def}$ 
correlations give us two independent ways to estimate BH masses in core Es.  

\end{abstract}


\section{Introduction}

\pretolerance=15000  \tolerance=15000

      Research on supermassive black holes (BHs) in galaxy nuclei centers
on the growing evidence that BH growth and galaxy evolution are connected (Ho 2004). 
The most striking evidence is the discovery of correlations between BH mass $M_\bullet$
and bulge (but not disk) luminosity 
(Kormendy 1993; 
Kormendy \& Richstone 1995; 
Magorrian \etal 1998;
Kormendy \& Gebhardt 2001)
and between $M_\bullet$ and the velocity dispersion $\sigma$ of the bulge outside the region 
influenced by the BH
(Ferrarese \& Merritt 2000;
Gebhardt \etal 2000b).
These scaling relations are clues to formation physics.~But they do not point 
to specific astrophysical processes.  So the literature on their interpretation has 
been extraordinarily wide-ranging.

     This paper reports tight mutual correlations between $M_\bullet$, $\sigma$, and the
light deficits $L_{\rm def}$ that define cores in elliptical galaxies.  These point more 
directly to formation processes.  In particular, we find a correlation between
the fundamental observables $L_{\rm def}$ and $\sigma$ that is as tight as the well 
known $M_\bullet$ -- $\sigma$ correlation.

      To set the stage, we recall that it is difficult to understand how cores can form in
galaxies that are made by major mergers.  We define\footnote{This is essentially equivalent 
to the classical definition (King 1978) that cores resemble the central profiles of 
non-singular isothermal spheres or the Nuker team definition that cores are regions near the center 
where the projected brightness profile shows a shallow cusp, $I(r) \propto r^{-\gamma}$ with 
$\gamma \leq 0.3$, interior to a downward profile break toward the center
(Kormendy \etal 1994;
Lauer \etal 1995;
Faber \etal 1997;
see Kormendy \etal 2008 for discussion).
}
a core as the region near the center of a bulge or elliptical interior to the radius where the 
outer, steep brightness profile shows a downward break to a shallow, central profile
(Kormendy 1999; 
Kormendy \etal 2008).
Caon \etal (1993) showed that the outer profiles $I(r)$  are best modeled by S\'ersic (1968) functions, 
$\log{I(r)} \propto r^{1/n}$.  Many authors confirm this.  Kormendy \etal (2008) find that 
S\'ersic functions fit ellipticals with remarkable precision over large dynamic ranges.  
With respect to these outer S\'ersic fits, cores show central light deficits.  Why?

      The problem is that galaxy mergers tend to preserve the highest progenitor densities.
Lower-luminosity galaxies have higher central densities 
(Kormendy 1985, 1987;
Lauer 1985;
Faber \etal 1997). 
When ellipticals or bulges merge, this tends to destroy the above correlation (Faber et al.\ 1997).   
A possible solution is that cores may be scoured by the orbital decay of binary BHs that form in galaxy mergers
Faber \etal 1997; 
Milosavljevi\'c \& Merritt 2001; 
Milosavljevi\'c et al.~2002; 
Merritt 2006).  The orbit shrinks as the binary flings stars away.  This decreases the surface 
brightness and excavates a core.  The effect of a series of mergers should be cumulative;
if the central mass deficit after one merger is $f M_\bullet$, then the mass deficit after $N$ major 
mergers should be $\sim N f M_\bullet$.  The above papers predict that $f \sim 0.5$ to 2.  Past
observations of mass deficits $M_{\rm def}$ were roughly consistent with this picture; $M_{\rm def}
\propto M_\bullet$ and $Nf \sim 1${\ts}--{\ts}5, consistent with formation by several successive 
dissipationless mergers (Milosavljevi\'c et al.~2001, 2002; Ravindranath \etal 2002;
Graham 2004; Ferrarese \etal 2006; Merritt 2006).

\section{Core Mass Deficit Versus BH Mass}

      The key to this paper is high-accuracy surface photometry of all known ellipticals in the 
Virgo cluster presented in Kormendy \etal (2008, hereafter KFCB).  Composite profiles are measured and
assembled from 6\ts--\ts11 sources for each core elliptical.
A variety of telescopes including the {\it Hubble Space Telescope\/} (HST) and wide-field, ground-based 
telescopes provide large dynamic range.  Comparison of many data sources minimizes systematic errors.
KFCB find that S\'ersic functions accurately fit \hbox{93\ts--\ts99\ts\%} of the light of each galaxy. 
Central departures of the profiles from these fits can be interpreted with confidence.
One result is an improved demonstration that profile shape participates in the well known dicotomy into 
two kinds of ellipticals 
(Bender 1988;
Bender et al.~1989; 
Nieto, Bender, \& Surma 1991; 
Kormendy \& Bender 1996).
HST photometry shows that ellipticals come in core and coreless varieties
(Lauer et{\ts}al.{\ts}\ts1995, 2005, 2007;
Gebhardt et{\ts}al.\ts1996; 
Rest~et{\ts}al.{\ts}2001;
Ravindranath \etal 2001).
Faber \etal (1997) show that this distinction correlates with the rest of the dichotomy:
core ellipticals generally have boxy-distorted isophotes and rotate slowly, while coreless
ellipticals have disky-distorted isophotes and rotate rapidly. KFCB confirm results of
Kormendy (1999) that coreless Es do not have featureless ``power law profiles'' near their 
centers but rather show distinct components -- extra light above the inward extrapolation of their
outer S\'ersic profiles.  The extra light resembles the extra central components predicted in
$n$-body simulations of galaxy mergers in which gas dissipation feeds starbursts (Mihos \& Hernquist 1994; 
see Hopkins \etal 2008 for extensive simulations).~KFCB conclude, in agreement with Faber \etal
(1997), that ``power law Es'' (now more accurately called ``extra light Es'') formed in dissipative (``wet'')
mergers with starbursts, whereas core Es formed in dissipationless (``dry'') mergers with binary BH scouring.

      KFCB calculate with improved accuracy the mass excesses in extra light and the mass deficits (not)
in cores.  Figure 1 shows that extra light correlates only weakly with $M_\bullet$. But mass deficits correlate 
closely with $M_\bullet$.  Moreover, the mass deficit is a larger multiple of the BH mass than previously thought.
The mean $<$\null$\log {M_{\rm def}/M_\bullet}$\null$>$ = $1.02 \pm 0.07$.   These values are surprisingly large 
in comparison to the prediction (Merritt 2006) that $M_{\rm def}/M_\bullet \simeq 0.5$ per major merger.  
However, with a more accurate treatment of the late stages of BH mergers, Merritt, Mikkola, \& Szell (2007)
find that $M_{\rm def}/M_\bullet$ can be as large as $\sim 4$ per merger.  Second, an additional process 
has been proposed to make large-$M_{\rm def}/M_\bullet$ cores (Merritt \etal 2004; Boylan-Kolchin, Ma, \& 
Quataert 2004; Gualandris \& Merritt 2008).  Coalescing binary BHs emit gravitational radiation anisotropically;
they recoil at velocities comparable to galaxy escape velocities.  If they do not escape, they decay back to the 
center by dynamical friction.  In the process, they heat the core.  Gualandris \& Merritt (2008) 
estimate that they can excavate as much as $M_{\rm def}/M_\bullet \sim 5$ in addition to the mass that 
was scoured by the pre-coalescence binary.  So our observations present no problem for the idea that cores 
are made by a combination of the above BH scouring mechanisms acting over the course
of one or more successive dry mergers.  We do not estimate the number of mergers, because theoretical
predictions of $f$ are uncertain and because our measurements of $M_{\rm def}$ depend on the assumption 
that unscoured profiles are exactly S\'ersic.

      The purpose of this {\it Letter\/} is to pursue the correlation in the bottom panel of Figure 1
in more detail.  In particular, we ``deconstruct'' it to correlations between basic observables.

      First, an explanation of the measurements is necessary.   The amount of extra or missing
light is calculated by integrating the two-dimensional brightness distribution of the galaxy 
from the center to the inner limit $r_{\rm min}$ of the outer S\'ersic function's radial fit range.  From this 
luminosity, we subtract the integral of the fitted S\'ersic function over the same radial range, keeping
the  ellipticity $\epsilon$ of the S\'ersic function fixed at $\epsilon(r_{\rm min})$.  Internal errors
are estimated by substituting plausible extrapolations of the $\epsilon(r)$ profile into the region at 
$r < r_{\rm min}$.  Our measurements depend critically on the high accuracy and large dynamic range of
the composite brightness profiles in KFCB; these constrain S\'ersic $n$ values better than previous
work.  Our values of $n$ are slightly larger than those derived previously (see KFCB), resulting
in larger estimates of light deficits.  Also, we calculate the difference between the integral of our outer
S\'ersic fit and the integral of the galaxy profile, not the difference between a S\'ersic fit and a 
core-S\'ersic fit as in Ferrarese \etal (2006).

\phantom{0}

\phantom{0}

\vskip 3.9truein
 
\includegraphics{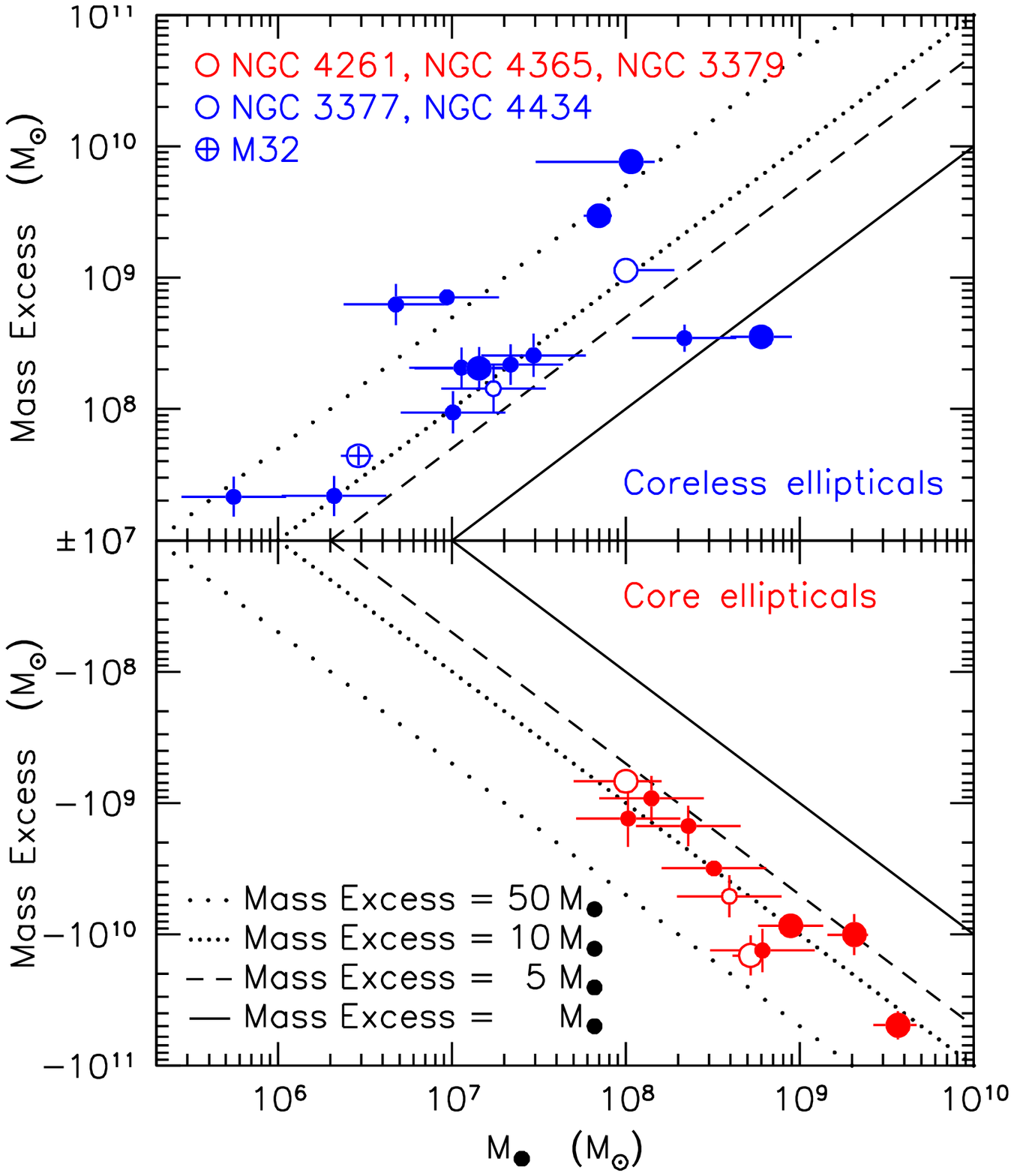}

\figcaption[]
{Stellar mass missing in cores ({\it lower panel}) or extra
in coreless galaxies ({\it upper panel\/}) as a function of BH mass. 
Large and small symbols are for galaxies with and without dynamical BH detections.
The sample is all known ellipticals in the Virgo cluster plus non-members
denoted by open symbols.  From Kormendy \etal (2008) with NGC 3379 added 
($\log M_\bullet = 8.0$).  
\lineskip=-4pt \lineskiplimit=-4pt
}

\vskip 15pt

\noindent Missing light is converted to missing mass using mass-to-light
ratios $M/L_V \propto L_V^{0.36}$ fitted to the SAURON sample of Cappellari \etal (2006) (zeropoint 
$M/L_V = 6.07$ at $M_V = -21.6$; estimated error in $\log {M/L_V} = 0.153$).  For galaxies with 
BH detections, we take $M/L_V$ and $M_\bullet$ from the dynamical modeling paper.  Otherwise, 
$M_\bullet$ is derived using the \hbox{$M_\bullet$\ts--\ts$\sigma$} correlation as fitted by 
Tremaine \etal (2002).  Velocity dispersions are mostly from Tremaine \etal (2002), Bender, 
Saglia, \& Gerhard (1994), and Kormendy \& Gebhardt (2001).  The sample, total absolute magnitudes 
$M_{VT}$, and light deficits $M_{V,\rm def}$ are from KFCB with NGC 3379 added.  Distances are
from Mei \etal (2007) or Tonry \etal (2001).  The data are listed in Table 1.

\includegraphics{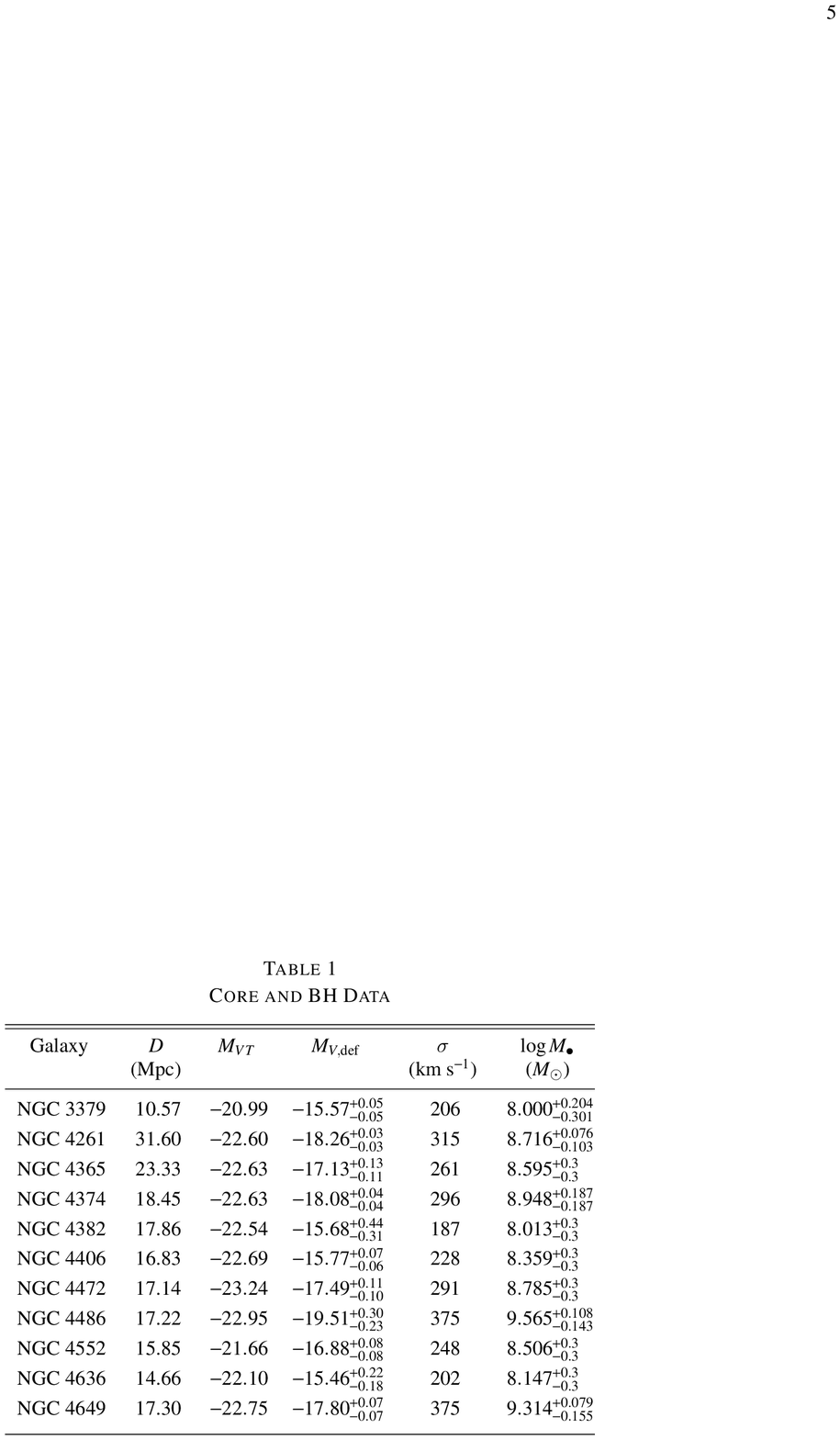}

\vfill\eject

\centerline{\null}

\vskip 3.0truein
 
\includegraphics{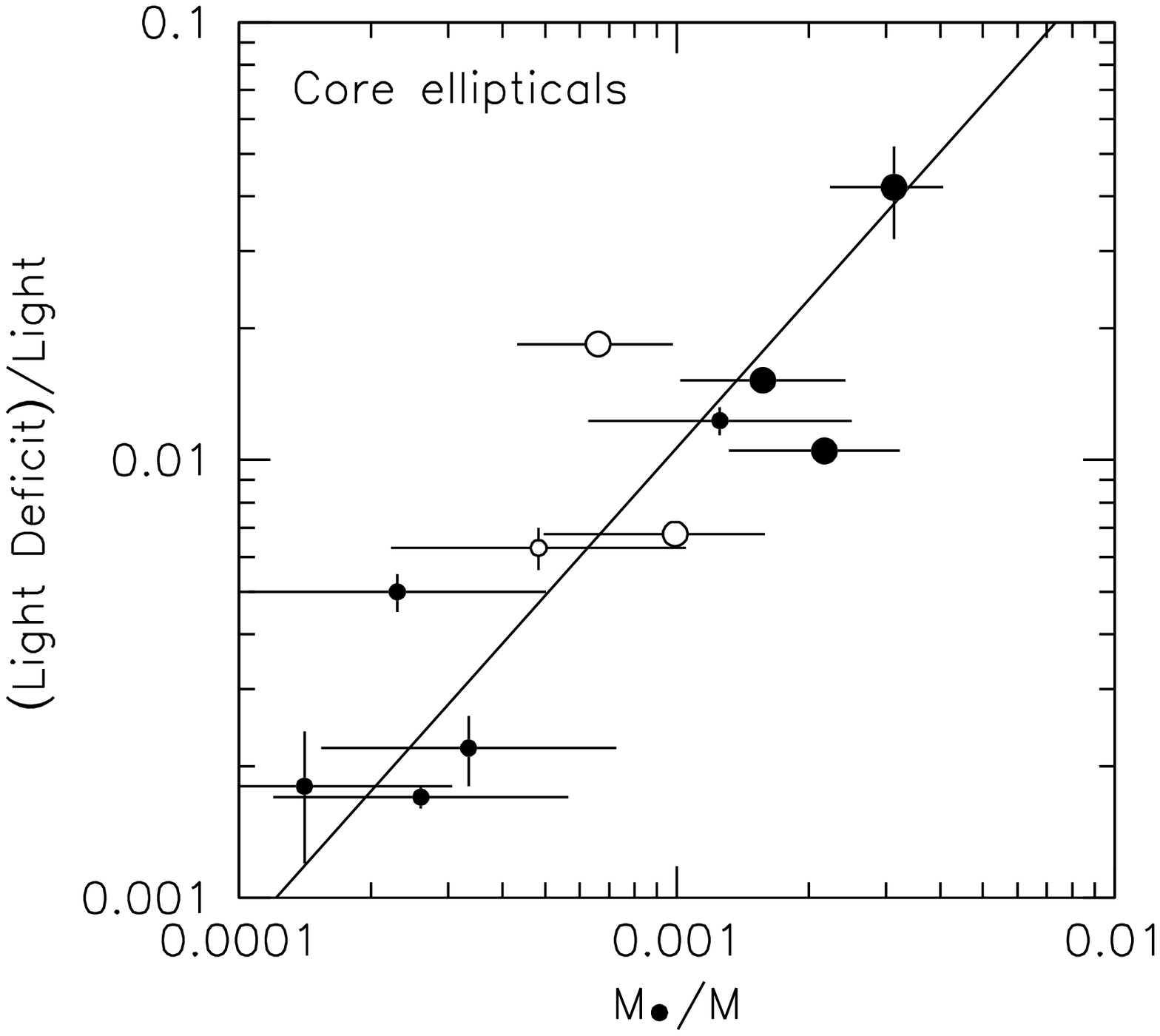}

\figcaption[]
{Fraction of the total $V$-band luminosity ``missing'' in cores versus the ratio of BH mass to the
total stellar mass of the galaxy.  The sample is as in Figure 1 and Table 1.  Large and small symbols 
denote galaxies with and without dynamical BH detections.   The galaxies with BH detections are 
({\it top to bottom\/}) 
M{\ts}87 (Macchetto \etal 1997), 
NGC 4261 (Ferrarese, Ford, \& Jaffe 1996), 
NGC 4374 (Bower \etal 1998),
NGC 4649 (Gebhardt \etal 2003), and
NGC 3379 (Gebhardt \etal 2000a).
The line is Equation (1).
\lineskip=-4pt \lineskiplimit=-4pt
}

\vskip 15pt

\section{Correlations Between Light Deficit and BH Mass}

      Figure 2 shows the fraction of the total $V$-band light of the galaxy that is ``missing'' in the
core versus the ratio of the BH  mass to the stellar mass of the galaxy. As we would expect if 
$M_{\rm def}/M_\bullet \simeq$ constant (Figure 1), there is a good correlation and the slope is 
consistent with 1.

      The lines in Figures 2 -- 4 are least-squares fits following Tremaine \etal (2002):
(i) the dependent and independent variables $x$ and $y$ are treated symmetrically; (ii) we calculate 
regressions of $y - \bar{y}$ on $x - \bar{x}$ so errors in the fit parameter are essentially uncorrelated 
and minimized (Tremaine \etal 2002, p.~742); (iii) we calculate a fit using the 
estimated errors in both parameters and adding intrinsic scatter to bring the reduced $\chi^2$ to 1; (iv)
we calculate a second fit by assigning equal errors to all $x$ and different equal errors to all $y$ and 
finding the values of those errors that result in a reduced $\chi^2$ of 1, thus providing partial protection 
against the danger that the assumed errors are not realistic, and (v) we adopt the mean of these fits as 
was done by Tremaine \etal (2002).  In the above, e.{\ts}g.,~$\bar{x}$ is the mean of the $x$ values.  In
Figure 2, the adopted correlation~is:
$$
{{L_{V,\rm def}/L_V}\over{0.01}} = \biggl(1.066^{+0.236}_{-0.193}\biggr)
                                   \biggl(  {  {M_\bullet/M}\over{0.001}  }  \biggr)^{1.123 \pm 0.235}, \eqno{(1)}
$$
\noindent with RMS scatter = 0.23 in $\log {L_{V,\rm def}/L_V}$ and 0.23 in $\log {M_\bullet/M}$.
The latter is formally less than the error $\pm 0.3$ in $\log M_\bullet$ adopted by Tremaine \etal
(2002).  The fraction of the light that is missing in cores may be as good a predictor of $M_\bullet$ 
as is $\sigma$.

      The canonical mean mass fraction in BHs is $M_\bullet/M \simeq 0.0013$ (Merritt \& Ferrarese 2001; 
Kormendy \& Gebhardt 2001).  The 5 BH detections in Figure 2 are consistent with this value.  But if
velocity dispersions reliably tell us BH masses, then there is significant scatter in $M_\bullet/M$
toward lower values.  Our observation that these lower $M_\bullet$ values are correlated with
lower $L_{\rm def}$ values adds confidence to our conclusion that $M_\bullet/M$ has significant
cosmic scatter.  The total range of $\log M_\bullet/M$ values found by Merritt \& Ferrarese (2001) and 
by Kormendy \& Gebhardt (2001) is more than $\pm 1$ around the mean.

      Figure 3 shows the absolute magnitude of the light that is missing in cores versus BH mass.
It is analogous to previous derivations (see references in \S\ts1) of the correlations of mass deficits
with BH masses, but $M_{V,\rm def}$ does not involve the uncertainty of mass-to-light ratios. 

\centerline{\null}

\vskip 3.15truein
 
\includegraphics{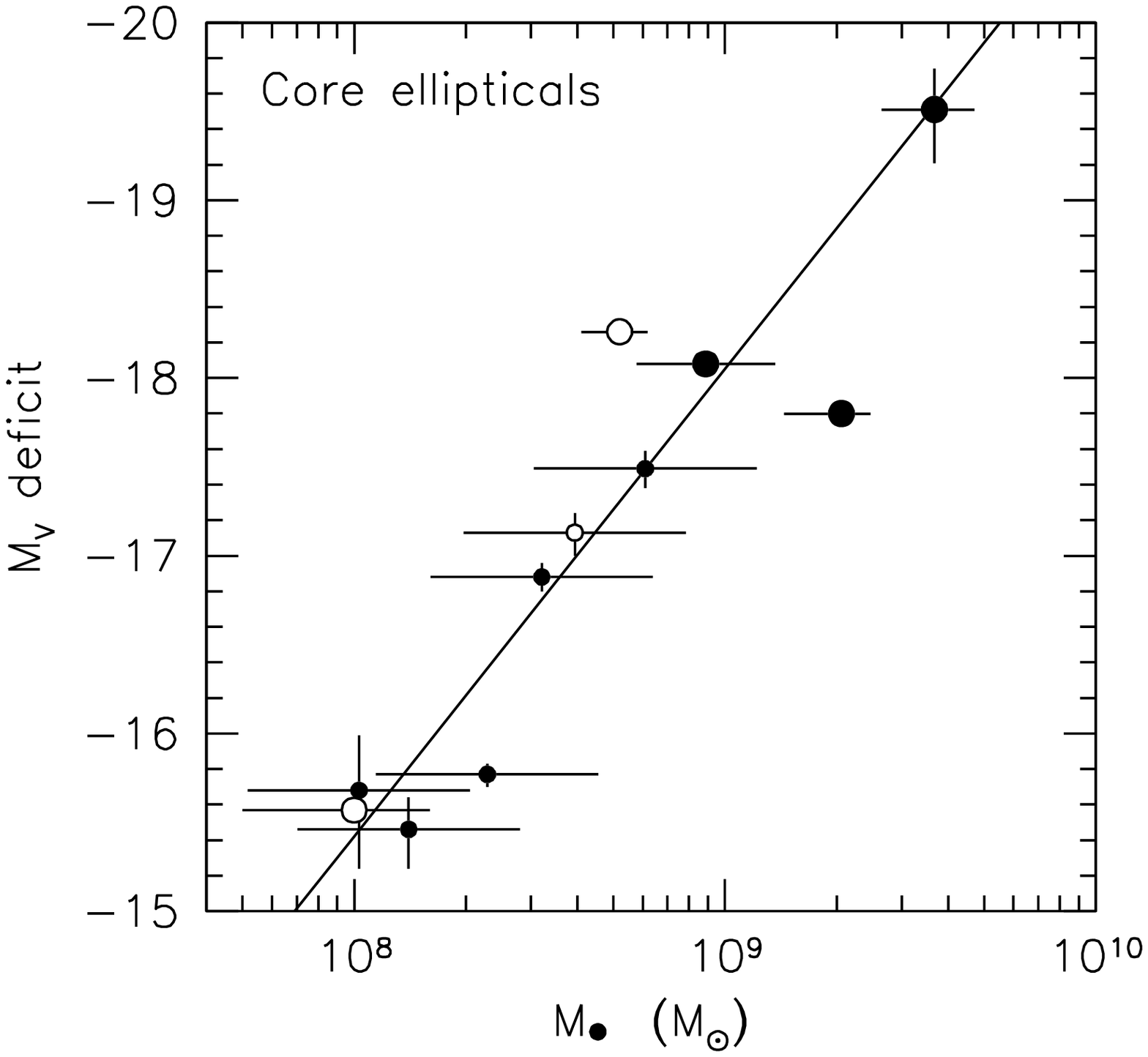}

\figcaption[]
{Absolute magnitude of the light that is missing in cores versus BH mass.
The sample and symbols are as in Figure 2.   The line is Equation (2).
\lineskip=-4pt \lineskiplimit=-4pt
}

\vskip 15pt
  
     The correlation in Figure 3 is remarkably good.  The fit is:
$$
M_{V,\rm def} = (-18.05 \pm 0.22) - (2.63 \pm 0.42)\ \log \biggl({M_\bullet \over {10^9~M_\odot}}\biggr), \eqno{(2)}
$$
\noindent with RMS scatter = 0.52 in $M_{V,\rm def}$ and 0.20 in $\log {M_\bullet}$.
The latter is less than the error $\pm 0.3$ in $\log M_\bullet$ taken from Tremaine \etal
(2002).  Again, core light deficit may be as good a predictor of $M_\bullet$ as is $\sigma$. 
Figure 3 spans the range of dynamically measured BH masses observed to date in ellipticals that have cores.

      For convenience, we can rewrite Equation (2) as:
$$
{M_\bullet \over {10^8~M_\odot}} = \biggl(0.81^{+0.18}_{-0.15}\biggr)\biggl({L_{V,\rm def} \over 
                                                                       {10^8~L_{V\odot}}}\biggr)^{0.95^{+0.18}_{-0.13}}.
\eqno{(3)}
$$
Remarkably, the typical $M/L_V \sim 8$ almost cancels the factor-of-10 ratio between $L_{V,\rm def}/L_V$ 
and $M_\bullet/M$ in Equation (1).  Within the errors, $M_\bullet \simeq L_{V,\rm def}$, where both are 
measured in Solar units.

      We emphasize that the $M_{V,\rm def}$ measurements in Figure 3 are independent of
$M_\bullet$ except for the indirect dependence of $M_\bullet$ on the light distribution through 
the dynamical modeling.  It is interesting that
the scatter in Figure 3, which mainly involves central galaxy quantities, is marginally smaller
than the scatter in Figure 2, which also involves the total galaxy luminosity.

\section{Light Deficit and Galaxy Velocity Dispersion}

      The observation that $M_\bullet$ correlates closely with both $\sigma$ and $M_{V,\rm def}$ hints
that there may be a good correlation between $M_{V,\rm def}$ and velocity dispersion.  Figure 4 shows that 
this correlation is, in fact, excellent.  Moreover, it is a correlation between pure observables, 
with no dynamical modeling involved.  

\centerline{\null}

\vskip 2.95truein
 
\includegraphics{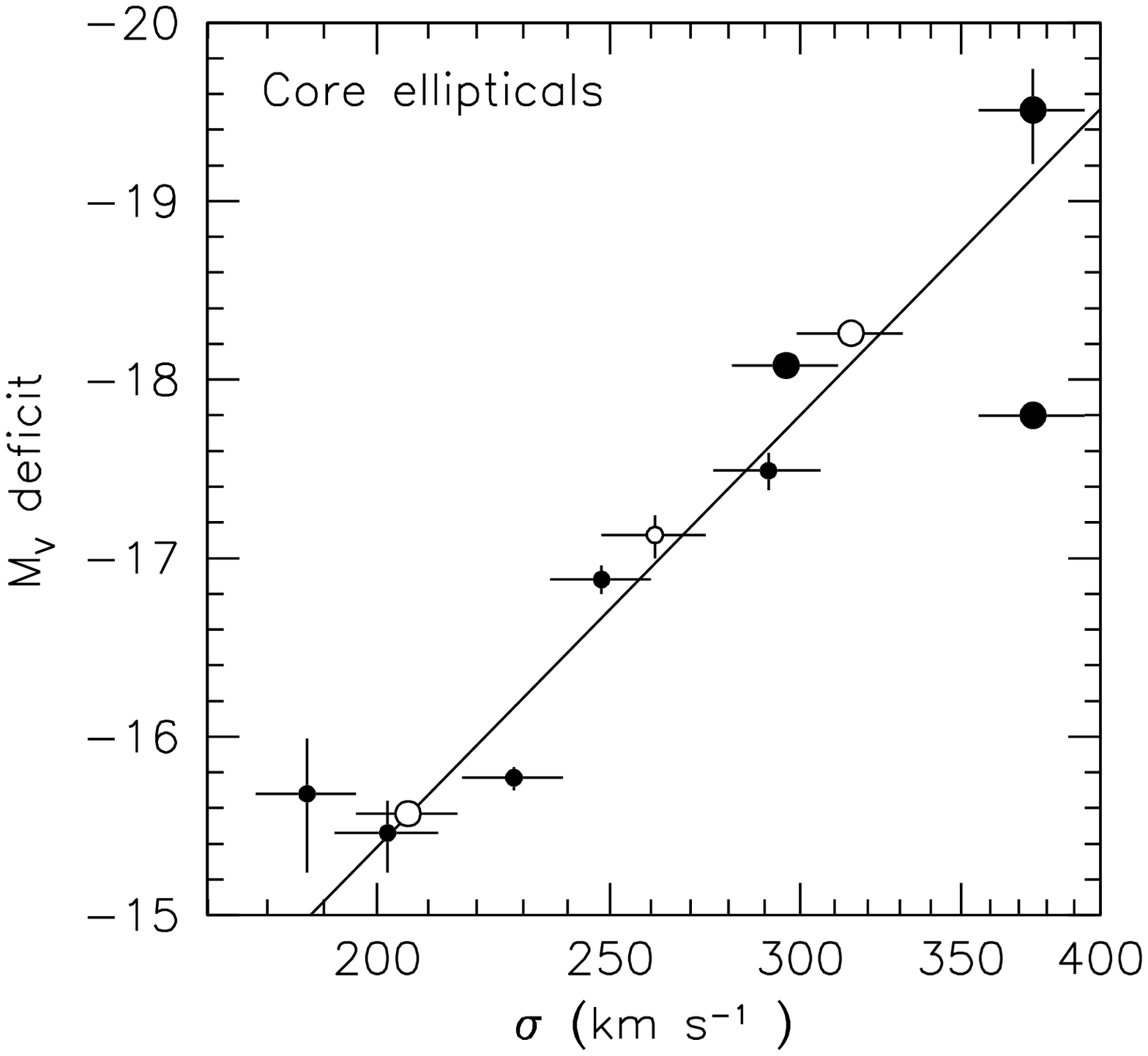}

\figcaption[]
{Absolute magnitude of the light missing in cores versus galaxy velocity dispersion averaged 
inside a slit aperture whose length is approximately twice the half-light radius (Table 1).  
\lineskip=-4pt \lineskiplimit=-4pt
}

\vskip 15pt

     The correlation in Figure 4 is:
$$
M_{V,\rm def} = (-16.71 \pm 0.18) - (13.73 \pm 1.92)\ \log \biggl({\sigma \over {250~{\rm km~s}^{-1}}}\biggr); \eqno{(4)}
$$
$$
{L_{V,\rm def} \over {10^8~L_{V\odot}}} = \biggl(4.10^{+0.75}_{-0.63}\biggr)\ 
                                           \biggl({{\sigma} \over 
                                                  {250~{\rm km~s}^{-1}}}\biggr)^{5.49 \pm 0.77}.
\eqno{(5)}
$$
\noindent The RMS scatter is 0.55 in $M_{V,\rm def}$, 0.22 in $\log L_{V,\rm def}$, and 0.0401 in $\log {\sigma}$.
Tremaine \etal (2002) estimate that the intrinsic dispersion in $M_\bullet$ at constant $\sigma$ is 
$0.23 \pm 0.05$ dex. Our observed scatter in $L_{V,\rm def}$ is essentially the same and includes
measurement errors.  For our sample, $\sigma$ predicts $L_{V,\rm def}$ slightly more accurately
than it predicts $M_\bullet$. % The correlation would be much tighter if NGC 4649 were omitted.

      Figures 2 -- 4 are remarkably good correlations, considering the difficulty of the $M_{V,\rm def}$
and $M_\bullet$ measurements.  The care that we took with the KFCB photometry helps.  But there was 
never a guarantee that the intrinsic scatter in such correlations would be small.  It is especially
surprising because only core galaxies participate.  Extra light galaxies contain BHs too (Figure 1), 
but extra light shows little correlation with $M_\bullet$.  KFCB suggest that core Es are made
in dry mergers whereas coreless Es are made in wet mergers in which central starbursts build the extra
light.  Then the amount of extra light is controlled mainly by the gas fraction in the progenitors and the 
properties of the starburst and only secondarily by the energy feedback from young stars, supernovae, and
nuclear activity.  We regard Figures 1\ts--\ts4 as further evidence for this explanation 
of the E{\ts}--{\ts}E dichotomy.

      In core galaxies, the excellent mutual correlations between light deficit, velocity dispersion, 
and BH mass -- with observed scatter roughly consistent with measurement errors -- imply that cores and
BHs are closely related.  We interpret them as a ``smoking gun'' in favor of the idea that cores are 
scoured by BHs.  Core formation was surprisingly regular, 
given the probably heterogeneous merger histories of elliptical galaxies.

      Our ``bottom line'' conclusion is that, for the present galaxy sample, $L_{V,\rm def}$ and $\sigma$ 
provide two independent and comparably accurate predictors of BH mass.

\acknowledgments

      It is a pleasure to thank David Fisher and Mark Cornell for an enjoyable collaboration
on KFCB.  JK thanks Milo\^s Milosavljevi\'c for helpful discussions on BH scouring.  We made extensive use 
of the NASA/IPAC Extragalactic Database (NED), which is operated by Caltech and JPL under contract with NASA, 
of the HyperLeda database http://leda.univ-lyon1.fr, and of NASA's Astrophysics Data System 
bibliographic services.  This work was supported by the National Science Foundation under grant AST-0607490.

\vfill\eject


\begin{references}

\reference{} Bender, R.~1988, A\&A, 193, L7  %  V anisotropy and isophote shape
\reference{} Bender, R., Saglia, R.~P., \& Gerhard, O.~E.~1994, MNRAS, 269, 785
\reference{} Bender, R., Surma, P., D\"obereiner, S., M\"ollenhoff, C., \& Madejsky,
             R. 1989, A\&A, 217, 35 
\reference{} Bower, G.~A., \etal 1998, ApJ, 492, L111
\reference{} Boylan-Kolchin, M., Ma, C.-P., \& Quataert, E.~2004, ApJ, 613, L37
\reference{} Caon, N., Capaccioli, M., \& D'Onofrio, M.~1993, MNRAS, 265, 1013
\reference{} Cappellari, M., \etal 2006, MNRAS, 366, 1126  
\reference{} Ebisuzaki, T., Makino, J., Okamura, S.~K.~1991, Nature, 354, 212
\reference{} Faber, S.~M., \etal 1997, AJ, 114, 1771
\reference{} Ferrarese, L., Ford, H.~C., \& Jaffe, W.~1996, ApJ, 470, 444
\reference{} Ferrarese, L., \& Merritt, D.~2000, ApJ, 539, L9
\reference{} Ferrarese, L., \etal 2006, ApJS, 164, 334
\reference{} Gebhardt, K., \etal 1996, AJ, 112, 105
\reference{} Gebhardt, K., \etal 2000a, AJ, 119, 1157
\reference{} Gebhardt, K., \etal 2000b, ApJ, 539, L13
\reference{} Gebhardt, K., \etal 2003, ApJ, 583, 92
\reference{} Graham, A.~W.~2004, ApJ, 613, L33
\reference{} Gualandris, A., \& Merritt, D.~2008, ApJ, 678, 780
\reference{} Ho, L.~C., Ed.~2004, Carnegie Observatories Astrophysics Series, Volume 1: Coevolution of 
             Black Holes and Galaxies (Cambridge University Press)
\reference{} Hopkins, P.~F., Cox, T.~J., Dutta, S.~N., Hernquist, L., Kormendy, J., \& Lauer, T.~R.~2008,
             ApJ, in press (arXiv:0805.3533)  %  Paper II
\reference{} King, I.~R.~1978, ApJ, 222, 1
\reference{} Kormendy, J.~1985, ApJ, 295, 73
\reference{} Kormendy, J.~1987, in IAU Symposium 127, Structure and Dynamics
             of Elliptical Galaxies, ed.~T.~de Zeeuw (Dordrecht: Reidel), 17
\reference{} Kormendy, J.~1993, in The Nearest Active Galaxies, ed.~J.~Beckman,
             L.~Colina \& H.~Netzer (Madrid: Consejo Superior de Investigaciones Cient\'\i ficas), 197
\reference{} Kormendy, J.~1999, in Galaxy Dynamics: A Rutgers Symposium, ed.~D.~Merritt,
             J.~A.~Sellwood, \& M.~Valluri (San Francisco: ASP), 124
\reference{} Kormendy, J., \& Bender, R.~1996, ApJ, 464, L119
\reference{} Kormendy, J., Fisher, D.~B., Cornell, M.~E., \& Bender, R.~2008, ApJS, in press (arXiv:0810.1681) (KFCB)
\reference{} Kormendy, J., \& Gebhardt, K.~2001, in 20$^{\rm th}$ Texas Symposium on 
             Relativistic Astrophysics, ed.~J.~C.~Wheeler \& H.~Martel (New York: AIP), 363
\reference{} Kormendy, J., \& Richstone, D.~1995, ARA\&A, 33, 581
\reference{} Kormendy, J., \etal 1994, in ESO/OHP Workshop on Dwarf Galaxies, 
             ed.~G. Meylan \& P.~Prugniel (Garching: ESO), 147
\reference{} Lauer, T.~R.~1985, ApJ, 292, 104
\reference{} Lauer, T.~R., \etal 1995, AJ, 110, 2622 % Nuker 1
\reference{} Lauer, T.~R., \etal 2005, AJ, 129, 2138
\reference{} Lauer, T.~R., \etal 2007, ApJ, 664, 226  %  Nuker 6 -- 12 authors
\reference{} Macchetto, F., Marconi, A., Axon, D.~J., Capetti, A., Sparks, W.,
             \& Crane, P.~1997, ApJ, 489, 579
\reference{} Magorrian, J., \etal 1998, AJ, 115, 2285  %  12 Au
\reference{} Mei, S., \etal 2007, ApJ, 655, 144  %  10 authors
\reference{} Merritt, D.~2006, ApJ, 648, 976
\reference{} Merritt, D., \& Ferrarese, L.~2001, MNRAS, 320, L30  %  MBH/Mbulge standard reference
\reference{} Merritt, D., Mikkola, S., \& Szell, A.~2007, ApJ, 671, 53
\reference{} Merritt, D., Milosavljevi\'c, M., Favata, M., Hughes, S.~A., \& Holz, D.~E.~2004,
             ApJ, 607, L9
\reference{} Mihos, J.~C., \& Hernquist, L.~1994, ApJ, 437, L47
\reference{} Milosavljevi\'c, M., \& Merritt, D.~2001, ApJ, ApJ, 563, 34
\reference{} Milosavljevi\'c, M., Merritt, D., Rest, A., \& van den Bosch, F.~C.~2002,
             MNRAS, 331, L51
\reference{} Nieto, J.-L., Bender, R., \& Surma, P.~1991, A\&A, 244, L37 
\reference{} Ravindranath, S., Ho, L.~C., \& Filippenko, A.~V.~2002, ApJ, 566, 801
\reference{} Ravindranath, S., Ho, L.~C., Peng, C.~Y., Filippenko, A.~V., \& Sargent,
             W.~L.~W.~2001, AJ, 122, 653
\reference{} Rest, A., \etal 2001, AJ, 121, 2431
\reference{} S\'ersic, J.~L.~1968, Atlas de Galaxias Australes (Cordoba: 
             Observatorio Astronomico, Universidad de Cordoba)
\reference{} Tonry, J.~L., \etal 2001, ApJ, 546, 681
\reference{} Tremaine, S., \etal 2002, ApJ, 574, 740

\end{references}
\end{document}